\documentclass[aps,amsmath,amssymb,showpacs]{revtex4}
\usepackage{graphicx}

\begin{document}
\title{Casimir-Lifshitz Interaction between Dielectric Heterostructures}

\author{Arash Azari, Himadri S. Samanta, and Ramin Golestanian}
\affiliation{Department of Physics and Astronomy, University of
Sheffield, Sheffield S3 7RH, United Kingdom}

\date{\today}

\begin{abstract}
The interaction between arbitrary dielectric heterostructures is
studied within the framework of a recently developed dielectric
contrast perturbation theory. It is shown that periodically
patterned dielectric or metallic structures lead to oscillatory
lateral Casimir-Lifshitz forces, as well as modulations in the
normal force as they are displaced with respect to one another. The
strength of these oscillatory contributions increases with
decreasing gap size and increasing contrast in the dielectric
properties of the materials used in the heterostructures.
\end{abstract}


\maketitle
\section{Introduction}

In light of the ongoing miniaturization of mechanical devices and
the recent developments in Casimir-Lifshitz interactions
\cite{Casimir48,Lifshitz,measure,lateral-exp,trench},  there has
been some recent interest in the effect of these interactions
between the components of small mechanical devices \cite{nanomech}.
Since these interaction are particularly strong at small distances,
it will be interesting to know how they can be utilized for
designing novel mechanical systems that could work without physical
contact and could potentially help solve the wear problem
\cite{machine}.

In the past few years there have been a surge of interest in
developing techniques that can be used to study the Casimir-Lifshitz
interaction in  non-ideal geometries, including geometry
perturbation theories \cite{GK,EHGK,lambrecht}, semiclassical
\cite{semiclass} and classical ray optics \cite{Jaffe}
approximations, multiple scattering and multipole expansions
\cite{balian,klich,multipole1,multipole2,multipole3}, world-line
method \cite{gies} and exact numerical diagonalization methods
\cite{Emig-exact,valery}, as well as the numerical Green's function
calculation method \cite{Johnson}. These methods have been used in
studying the Casimir force in a variety of different geometries,
which have improved significantly our understanding of the
nontrivial geometry dependence of this effect.

\begin{figure}
\includegraphics[width=0.4\columnwidth]{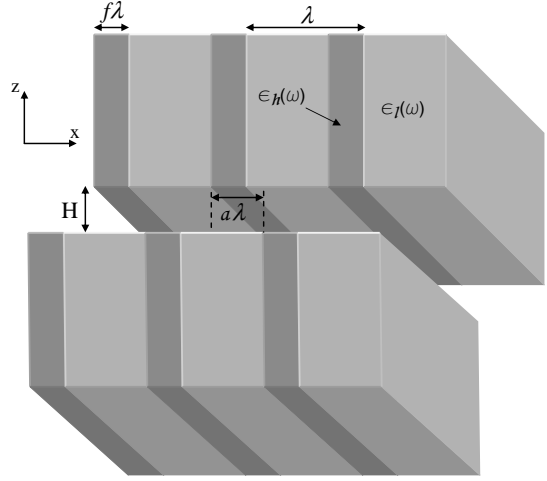}
\caption{Schematic representation of two identical semi-infinite and
periodic objects made of intercalated layers of {\em high} and {\em
low} dielectric functions, occupying the fractions of $f$ and $1-f$,
respectively. Here $H$ is the separation between them, $a$ is a
dimensionless lateral displacement, and $\lambda$ is the wavelength
of the periodic structure.} \label{fig:schem}
\end{figure}

The effect of non-ideal geometry has been shown to lead to a number
interesting effects. For example, it has been suggested that
corrugated surfaces opposite one another can experience an
oscillatory lateral Casimir force \cite{GK}, which was subsequently
observed experimentally \cite{lateral-exp}. A recent experiment
probing the normal Casimir force between a smooth surface and
surface with tall rectangular corrugations also revealed further
evidence on the non-additive nature of the Casimir force
\cite{trench}. Here, we study the Casimir-Lifshitz interaction
between arbitrary dielectric heterostructures within the framework
of a recently developed formalism \cite{ramin,rg-09}. We derive a
closed form expression for the Casimir-Lifshitz energy between two
dielectric heterostructures (such as the example depicted in Fig.
\ref{fig:schem}) up to the second order in the perturbation theory
and show that a coherent coupling between the different modes of the
spectrum of the dielectric pattern takes place across the gap. As a
special example, we consider unidirectional periodic
heterostructures (see Fig. \ref{fig:schem}) and calculate the
lateral and normal Casimir-Lifshitz force between them within the
same order in the perturbation theory. We find that coupling between
modes with identical wavevectors of the pattern structures between
the different objects can lead to modulations in the normal force
and can give rise to oscillatory later forces, reminiscent of the
lateral Casimir force that appears due to coupling between
geometrical features such as corrugations \cite{GK,lateral-exp}.

This paper is organized as follows. Section \ref{sec:form} sketches
the dielectric contrast perturbation theory, and Sec.
\ref{sec:diel-het} elaborates on how it can be used for dielectric
heterostructures giving closed form expressions for the second order
term in the perturbation theory. Section \ref{sec:norm-lat} gives
the results for the lateral and normal Casimir-Lifshitz force for a
number of choices of materials, and Sec. \ref{sec:disc} contains
some discussions and concluding remarks.

\section{Theoretical Formulation}\label{sec:form}

To calculate the Casimir-Lifshitz interaction we need to quantize
the electromagnetic field in a background that includes the
dielectric or metallic objects that modify the quantum fluctuations
of the field. Describing a general assortment of dielectric and
metallic objects in space via a frequency dependent dielectric
profile $\epsilon(\omega,{\bf r})$, we can write a general
expression for the Casimir-Lifshitz energy as \cite{rg-09}
\begin{equation}
E=\hbar \int_0^\infty \frac{d \zeta}{2 \pi}\; {\rm tr} \ln
\left[{\cal K}_{ij}(\zeta;{\bf r},{\bf r}')\right],\label{E-1}
\end{equation}
where
\begin{equation}
{\cal K}_{ij}=\left[\frac{\zeta^2}{c^2} \epsilon(i \zeta,{\bf
r})\delta_{ij}+\partial_i
\partial_j-\partial_k \partial_k \delta_{ij}\right] \delta^3({\bf
r}-{\bf r}'). \label{Kij-1}
\end{equation}
We can consider the dielectric function as $\epsilon(i \zeta,{\bf
r})=1+\delta \epsilon(i \zeta,{\bf r})$, and expand Eq. (\ref{E-1})
in powers of the dielectric contrast. A similar approach has been
the subject of a few recent studies
\cite{barton,ramin,buhmann,rudi,milton}.

The expansion leads to the decomposition of ${\cal K}_{ij}$ into a
diagonal part ${\cal K}_{0,ij}$, corresponding to the empty space,
and a perturbation part $\delta {\cal K}_{ij}$, namely
\begin{equation}
{\cal K}_{ij}(\zeta;{\bf q},{\bf q}')={\cal K}_{0,ij}(\zeta,{\bf q})
(2 \pi)^3 \delta^3({\bf q}+{\bf q}')+\delta {\cal K}_{ij}(\zeta;{\bf
q},{\bf q}'),
\end{equation}
where
\begin{equation}
{\cal K}_{0,ij}(\zeta,{\bf q})=\frac{\zeta^2}{c^2} \delta_{ij}+q^2
\delta_{ij}-q_i q_j,
\end{equation}
and
\begin{equation}
\delta {\cal K}_{ij}(\zeta;{\bf q},{\bf q}')=\frac{\zeta^2}{c^2}
\delta_{ij} \delta \tilde{\epsilon}(i \zeta,{\bf q}+{\bf q}').
\end{equation}
This yields an expansion
\begin{equation}
{\rm tr} \ln [{\cal K}]={\rm tr} \ln [{\cal
K}_0]+\sum_{n=1}^{\infty} \frac{(-1)^{n-1}}{n} \; {\rm tr}[({\cal
K}_0^{-1} \delta {\cal K})^n],\label{tr-eq}
\end{equation}
where
\begin{equation}\displaystyle
{\cal K}_{0,ij}^{-1}(\zeta,{\bf q})=\frac{\frac{\zeta^2}{c^2}
\delta_{ij}+q_i q_j}{\frac{\zeta^2}{c^2}
\left[\frac{\zeta^2}{c^2}+q^2\right]}.
\end{equation}
The first term is the vacuum energy in the absence of the objects,
and the terms in the series take account of their effect in a
perturbative scheme. The $n$-th order term in Eq. (\ref{tr-eq})
takes on the explicit form
\begin{equation}
{\rm tr}[({\cal K}_0^{-1} \delta {\cal K})^n]=\int \frac{d^3 {\bf
q}^{(1)}}{(2 \pi)^3} \cdots \frac{d^3 {\bf q}^{(n)}}{(2 \pi)^3}
\frac{[\frac{\zeta^2}{c^2} \delta_{i_{1}i_{2}}+q_{i_{1}}^{(1)}
q_{i_{2}}^{(1)}] \cdots [\frac{\zeta^2}{c^2}\delta_{i_{n}
i_{1}}+q_{i_{n}}^{(n)}
q_{i_{1}}^{(n)}]}{[\frac{\zeta^2}{c^2}+q^{(1)2}] \cdots
[\frac{\zeta^2}{c^2} +q^{(n)2}]} \delta \tilde{\epsilon}(i
\zeta,-{\bf q}^{(1)}+{\bf q}^{(2)}) \cdots \delta \tilde{\epsilon}(i
\zeta,-{\bf q}^{(n)}+{\bf q}^{(1)}),\label{higher}
\end{equation}
which involves the Fourier transform of the dielectric contrast
profile. Going to real space, we can rewrite the energy of the
system as \cite{rg-09}
\begin{equation}
E=\hbar \int_0^\infty \frac{d \zeta}{2 \pi} \;\sum_{n=1}^{\infty}
\frac{(-1)^{n-1}}{n} \int d^3 {\bf r}_1 \cdots d^3 {\bf r}_n {\cal
A}_{i_{1}i_{2}}({\bf r}_1-{\bf r}_2) \cdots {\cal
A}_{i_{n}i_{1}}({\bf r}_n-{\bf r}_1) \left[\frac{\delta \epsilon(i
\zeta,{\bf r}_1)}{1+\frac{1}{3} \delta \epsilon(i \zeta,{\bf r}_1)
}\right] \cdots  \left[\frac{\delta \epsilon(i \zeta,{\bf
r}_n)}{1+\frac{1}{3} \delta \epsilon(i \zeta,{\bf r}_n) }\right],
\label{E-gen-2}
\end{equation}
where
\begin{equation}
{\cal A}_{ij}(\zeta,{\bf r})=\frac{\zeta^2}{c^2} \frac{{\rm
e}^{-\zeta r/c}}{4 \pi r} \left[\delta_{ij} \left(1+\frac{c}{\zeta
r}+\frac{c^2}{\zeta^2 r^2}\right)-\frac{r_i r_j}{r^2} \left(1+3
\frac{c}{\zeta r}+3 \frac{c^2}{\zeta^2 r^2}\right)\right],
\label{eq:Ar-def}
\end{equation}
We now use this formulation to study the Casimir-Lifshitz
interaction between structures with inhomogeneous or patterned
dielectric properties.

\section{Dielectric Heterostructures} \label{sec:diel-het}

Let us now consider a configuration similar to the one depicted in
Fig. \ref{fig:schem}, namely two dielectric heterostructures that
are placed parallel to each other at a separation $H$. Using the
definition ${\bf r}=({\bf x},z)$, the dielectric profile can be
written as
\begin{equation}
\epsilon(i \zeta,{\bf r})=\left\{\begin{array}{ll}
\epsilon_u(i \zeta,{\bf x}), & \;\;\;\; \frac{H}{2} \leq z < +\infty,   \\\\
1, & \;\;\;\; \frac{-H}{2} < z < \frac{H}{2},   \\\\
\epsilon_d(i \zeta,{\bf x}), & \;\;\;\; -\infty < z \leq
\frac{-H}{2},
\end{array} \right. \label{epsilon-profile}
\end{equation}
using the labels {\em u} and {\em d} for the ``up'' and ``down''
bodies respectively.

To keep the calculations tractable, we now focus on the second order
term in the series expansion in Eq. (\ref{E-gen-2}). For such two
semi-infinite bodies, the second order interaction term between the
bodies can be written as
\begin{equation}
E_2=-\frac{\hbar}{2 \pi^2 c^2} \int_0^\infty d \zeta \; \zeta^2 \int
d^2{\bf x} d^2{\bf x}' \int \frac{d^2 {\bf Q}}{(2 \pi)^2} \;{\rm
e}^{i {\bf Q} \cdot ({\bf x}-{\bf x}')} \;{\cal
E}(Q)\left[\frac{\delta \epsilon_u(i \zeta,{\bf x})}{1+\frac{1}{3}
\delta \epsilon_u(i \zeta,{\bf x})}\right]\left[\frac{\delta
\epsilon_d(i \zeta,{\bf x}')}{1+\frac{1}{3} \delta \epsilon_d(i
\zeta,{\bf x}')}\right],\label{E2}
\end{equation}
for any lateral dielectric function profile, where
\begin{equation}
{\cal E}(Q)=\int_{1}^{\infty}d p \;\frac{[2p^{4}-2p^{2}+1]}{\left[4
p^{2}+(c Q/\zeta )^{2}\right]^{3/2}}\; {\rm e}^{-\frac{\zeta
H}{c}\sqrt{4 p^{2}+(c Q/\zeta)^{2}}},\label{EQ-def}
\end{equation}
and $\delta\epsilon_{u,d}( i \zeta,{\bf x})=\epsilon_{u,d}(i
\zeta,{\bf x})-1$.

We now focus on the specific example of unidirectional periodic
structures as depicted in Fig. \ref{fig:schem}, which is made of
subsequent layers of materials with relatively {\em high} and {\em
low} dielectric functions. We can use the periodic properties of the
dielectrics and write them in Fourier series expansion.  As Fig.
\ref{fig:schem} shows, we can define the dielectric profile of the
$d$-object as
\begin{equation}
\epsilon_{d} \left(i\zeta,x \right)=\left\{
\begin{array}{ll}
\epsilon _{l}\left(i\zeta \right),  & \;\;\;\;
-\frac{\lambda}{2}+s\lambda \leq x \leq
-\frac{f\lambda}{2}+s\lambda,
\\\\
\epsilon _{h}\left(i\zeta \right),  & \;\;\;\; -\frac{f
\lambda}{2}+s\lambda < x < \frac{f\lambda}{2}+s\lambda,
\\\\
\epsilon _{l}\left(i\zeta \right),  & \;\;\;\; \frac{f
\lambda}{2}+s\lambda \leq x \leq \frac{\lambda}{2}+s\lambda,
\end{array}
\right.\label{diel-profile}
\end{equation}
where $s$ is an integer number. We define the Fourier series as
\begin{equation}
\frac{\delta\epsilon_{d} \left(i\zeta,x
\right)}{1+\frac{1}{3}\delta\epsilon_{d} \left(i\zeta,x
\right)}=\sum_{m=-\infty}^{\infty} {\mathcal C}_{m}(i \zeta) \;
{\rm e}^{i 2 \pi m x/\lambda},
\end{equation}
where
\begin{equation}
{\mathcal C}_{m}(i \zeta)=\frac{\sin m \pi f}{m \pi}
\left[\frac{\delta\epsilon_{h}
\left(i\zeta\right)}{1+\frac{1}{3}\delta\epsilon_{h}
\left(i\zeta\right)}-\frac{\delta\epsilon_{l}
\left(i\zeta\right)}{1+\frac{1}{3}\delta\epsilon_{l}
\left(i\zeta\right)}\right],
\end{equation}
for $m \neq 0$, and
\begin{equation}
{\mathcal C}_{0}(i \zeta)=f \left[\frac{\delta\epsilon_{h}
\left(i\zeta\right)}{1+\frac{1}{3}\delta\epsilon_{h}
\left(i\zeta\right)}\right]+(1-f) \left[\frac{\delta\epsilon_{l}
\left(i\zeta\right)}{1+\frac{1}{3}\delta\epsilon_{l}
\left(i\zeta\right)}\right].
\end{equation}
We can find the corresponding expansion for the $u$-object by
changing $x\rightarrow x+a\lambda$.

Using the Fourier series expansion, one can find the
Casimir-Lifshitz energy between two dielectric heterostructures as
depicted in Fig. \ref{fig:schem} [up to second order in the
Clausius-Mossotti expansion of Eq. (\ref{E-gen-2})] as
\begin{equation}
E_{pp}=-\frac{\hbar A}{2\pi^{2} c^2}{\sum_{m=0}^{\infty}}^{'}
\int_{0}^{\infty}d\zeta \;\zeta^2 {\cal E}\left(\frac{2 \pi
m}{\lambda}\right)\;{\mathcal C}^{2}_{m}(i\zeta) \cos (2 \pi m
a),\label{Epp}
\end{equation}
where the prime on the summation sign indicates that the $m=0$ term
is counted with half the weight, and the $pp$ index means the energy
calculated for the plate-plate geometry. This result shows that
similar to the case of two corrugated surfaces, two patterned
dielectric heterostructures also couple to each other at the leading
order when the two wavelengths of the modulations are equal
\cite{GK}. Moreover, higher harmonics contribute to the
Casimir-Lifshitz energy with exponentially decaying contributions,
such that at large separations only the fundamental mode (lowest
harmonic) will survive \cite{Emig-exact}.

\section{The Normal and Lateral Forces}\label{sec:norm-lat}

We now use Eq. (\ref{Epp}) to calculate the normal and lateral
forces between different types of dielectric and metallic
heterostructures. We look at three different types of materials as
examples, namely, gold, silicon, and air/vacuum, and consider
layered materials made of gold-silicon, silicon-air, and gold-air.
We describe the dielectric function of gold using a plasma model,
namely,
\begin{math} \displaystyle
\epsilon (i \zeta)=1+\frac{\omega_{p}^{2}}{\zeta^{2}},
\end{math}
where $\omega_{p}$ is the plasma frequency, which is given as
$\omega_{p}({\rm Au})=1.37\times10^{16}$ rad/s \cite{Palik}. For
silicon we use the Drude-Lorentz form
\begin{math} \displaystyle
\epsilon (i
\zeta)=1+\frac{\omega_{p}^{2}}{\zeta^{2}+\omega_{0}^{2}},
\end{math}
where $\omega_{p}({\rm Si})=3.3\; \omega_{0}({\rm Si})$ and
$\omega_{0}({\rm Si})=6.6\times 10^{15}$ rad/s \cite{Palik}.
Finally, for air/vacuum we use $\epsilon (i \zeta)=1$.

Due to difficulties in keeping the surfaces of the objects parallel
to each other, most experiments are performed in plate-sphere
geometry. To perform the calculation of the forces for the
plate-sphere configuration, we can use the Derjaguin Approximation
\cite{Israelachvili92}, where we replace one of the semi-infinite
objects with a planar surface with a sphere with radius $R$. The
approximation is valid provided that the radius of sphere is much
larger than the distance between the dielectric heterostructures,
namely, $R\gg H$. Using this approximation we can find the normal
force between a semi-infinite dielectric heterostructure and a
sphere of the same material composition as \cite{Israelachvili92}
\begin{equation}
F_{ps}^{nor}=2 \pi R \left(\frac{E_{pp}}{A}\right).\label{Fnorm}
\end{equation}
Using this result, we can find the Casimir-Lifshitz energy for
plate-sphere configuration as \begin{equation}
E_{ps}=-\int_{H}^{\infty}d H' \;F_{ps}^{nor}(H'),\label{Eps}
\end{equation}
which we can now use to calculate the lateral Casimir force as
\begin{equation}
F_{ps}^{lat}=- \frac{1}{\lambda} \;\frac{\partial E_{ps}}{\partial
a}.\label{Fps}
\end{equation}
Substituting Eqs. (\ref{Fnorm}) and (\ref{Eps}) into Eq.
(\ref{Fps}), it reads
\begin{equation}
F_{ps}^{lat}=\frac{2 \pi R}{\lambda} \;\frac{\partial}{\partial
a}\int_{H}^{\infty} d H' \; \left(\frac{E_{pp}(H')}{A}\right).
\end{equation}
The above equations are the basis of the results that will be
presented below.

\begin{figure}
\includegraphics[width=.3\columnwidth]{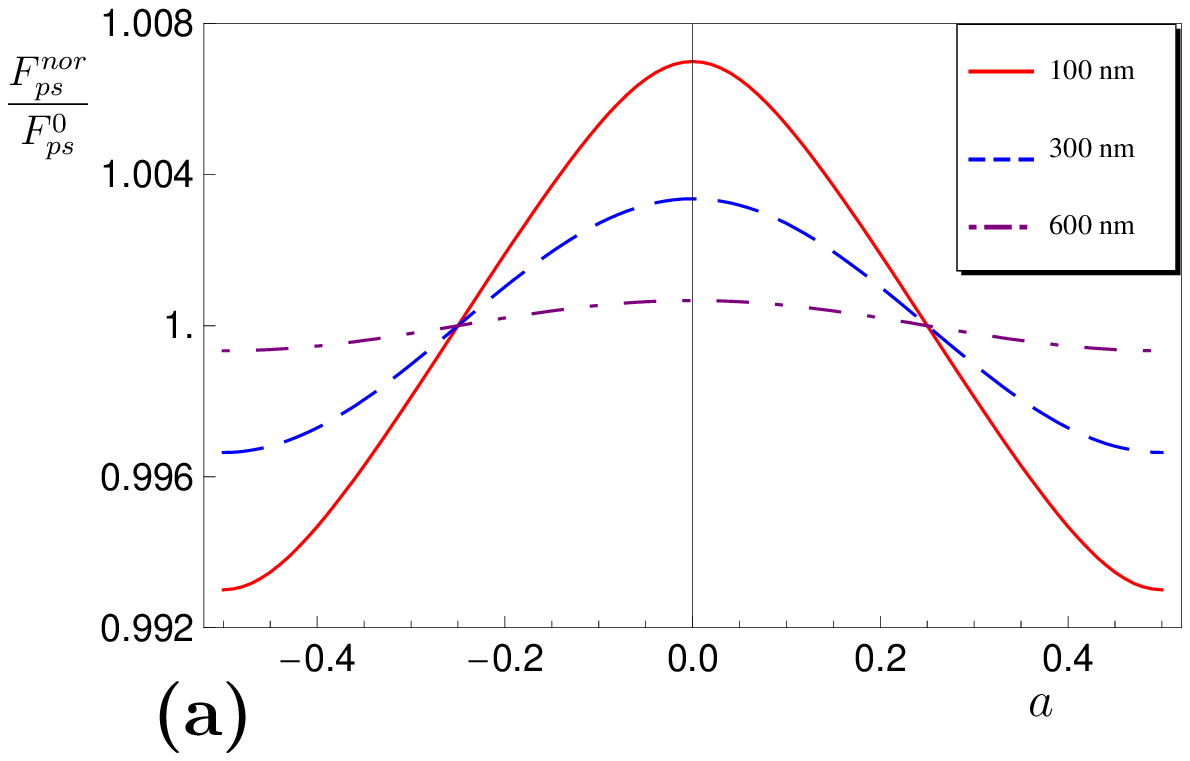}
\hspace{.5cm}
\includegraphics[width=.3\columnwidth]{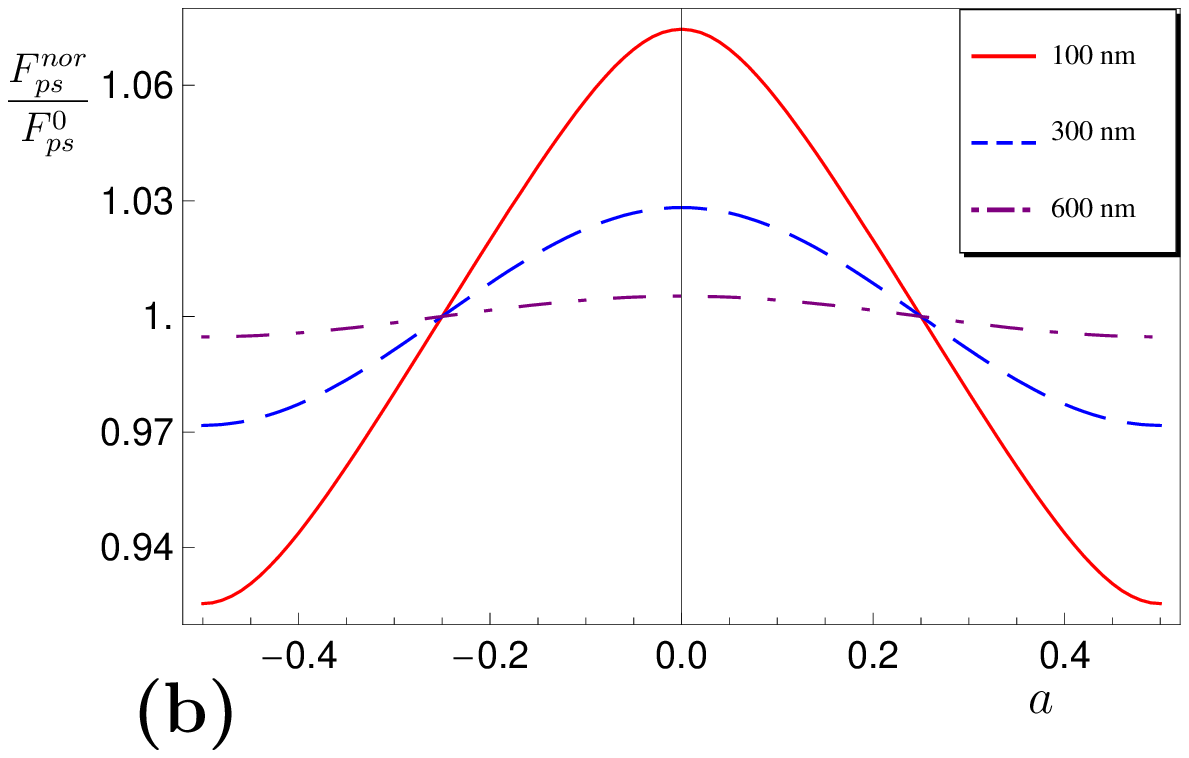}
\hspace{.5cm}
\includegraphics[width=.3\columnwidth]{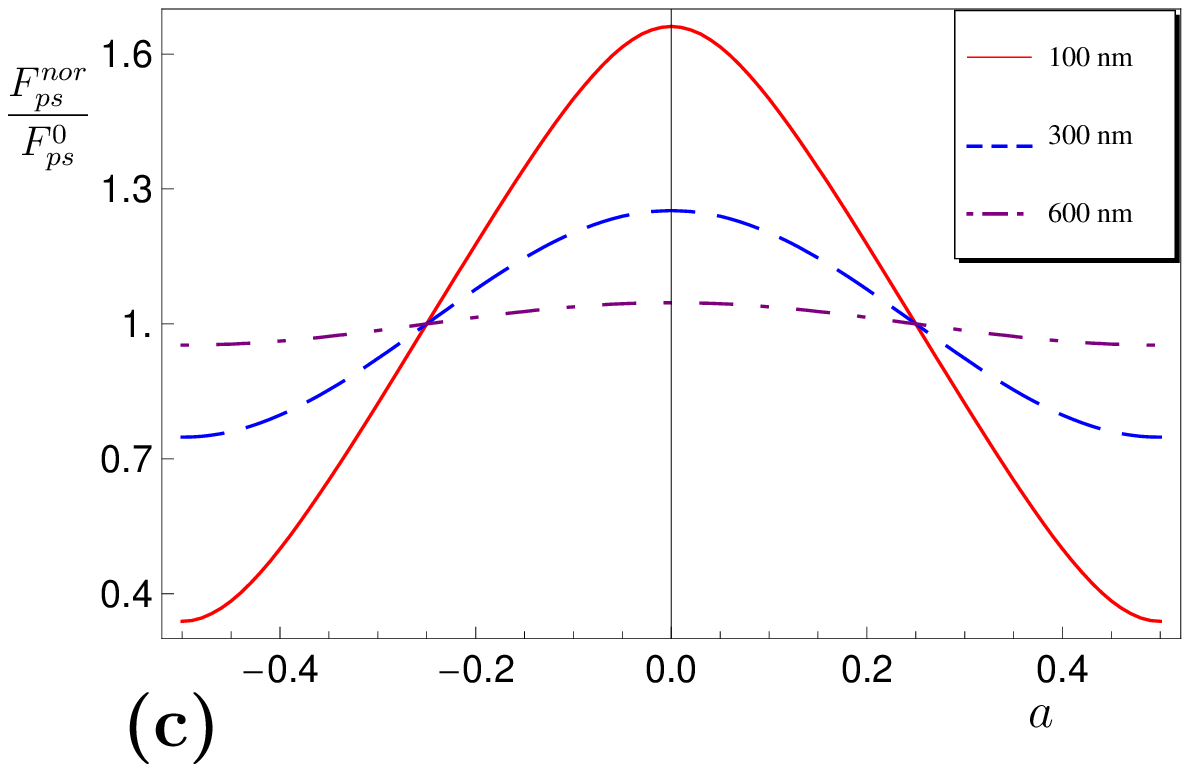}
\includegraphics[width=.3\columnwidth]{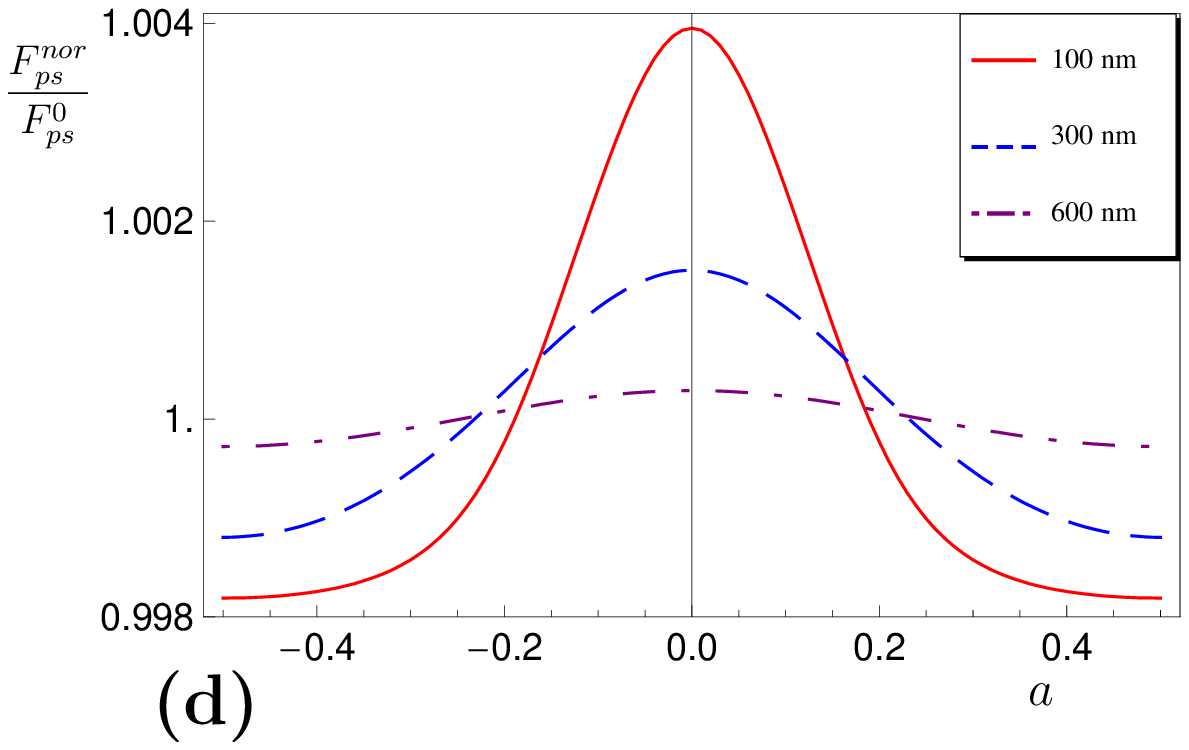}
\hspace{.5cm}
\includegraphics[width=.3\columnwidth]{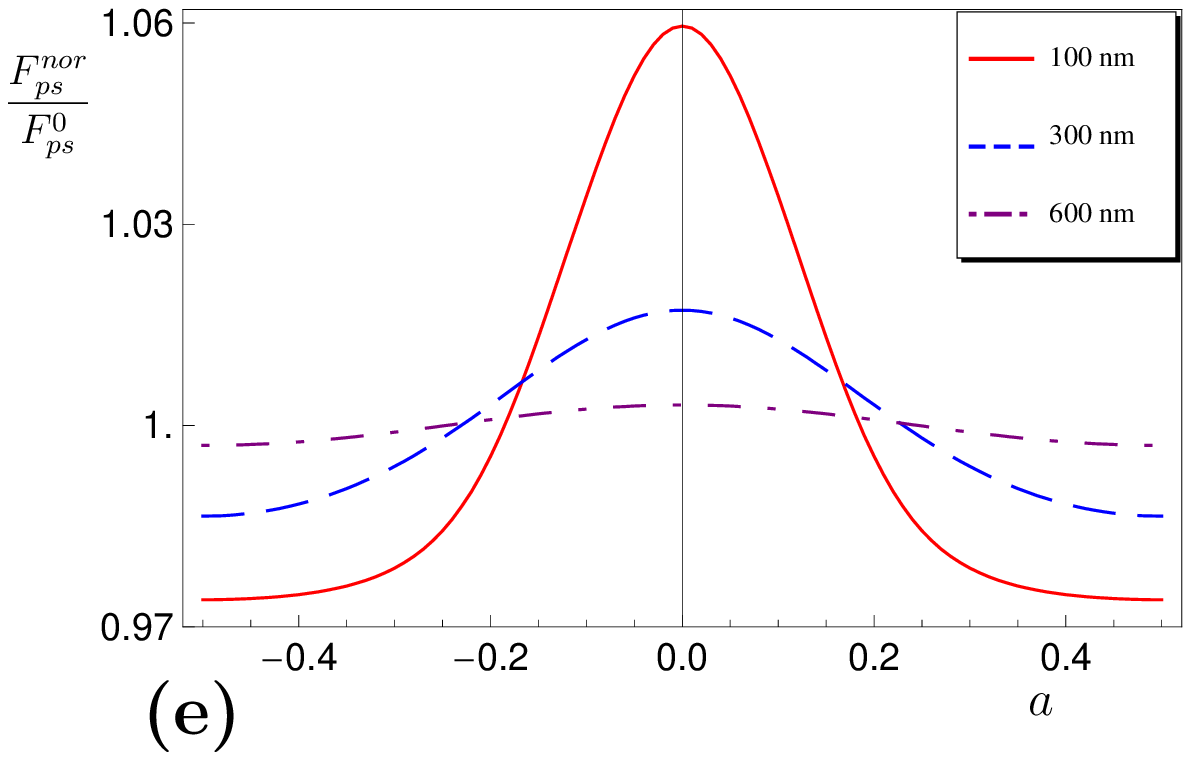}
\hspace{.5cm}
\includegraphics[width=.3\columnwidth]{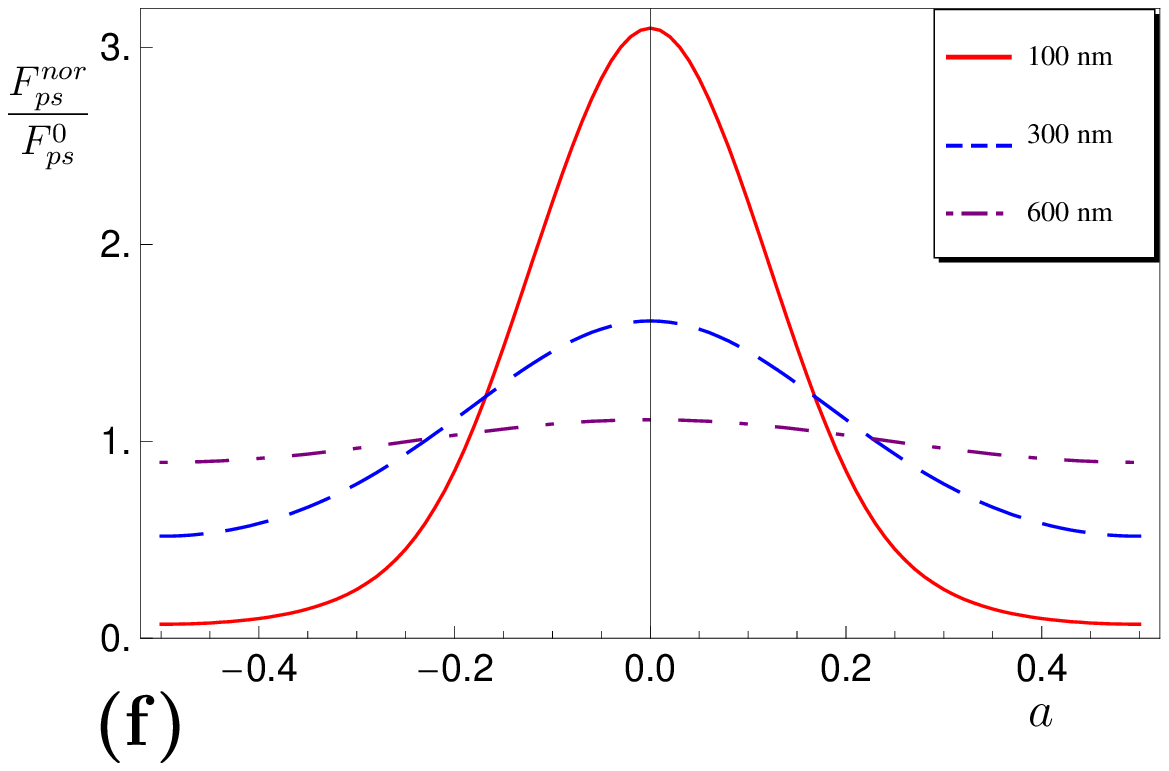}
\caption{Normal Casimir-Lifshitz force between layered dielectric
heterostructures as shown in Fig. \ref{fig:schem} in the
plate-sphere geometry, corresponding to (a) gold-silicon, (b)
silicon-air, and (c) gold-air, with $f=0.5$, and to (d)
gold-silicon, (e) silicon-air, and (f) gold-air, with $f=0.2$. The
numerical value of corrugation wavelength used is $\lambda=1$
$\mu$m. Different curves correspond to different gap sizes of
$H=100$ nm, $H=300$ nm, and $H=600$ nm. The forces are normalized to
$F_{ps}^0$ that corresponds to the normal force when the
laterally-averaged dielectric profile is used.} \label{fig:norm}
\end{figure}

Figures \ref{fig:norm}a-c show the normal Casimir-Lifshitz force
between two unidirectional (layered) dielectric heterostructures as
shown in Fig. \ref{fig:schem} when laterally displaced with respect
to one another by $a \lambda$. It corresponds to the symmetric case
with $f=0.5$, and the corrugation wavelength of $\lambda=1$ $\mu$m.
Three different compositions of gold-silicon, silicon-air, and
gold-air are considered each at three different gap sizes of $H=100$
nm, $H=300$ nm, and $H=600$ nm. The normal forces are normalized
using the normal force $F_{ps}^0$ that corresponds to the
Casimir-Lifshitz force calculated within the same scheme but with
laterally averaged dielectric profile, which corresponds to the
$m=0$ term in the expansion in Eq. (\ref{Epp}). The normal force is
found to oscillate as a function of the lateral displacement, having
the maximum value when the regions of high dielectric constant from
both sides are exactly opposite one another, and the minimum value
when in the staggered configuration where regions of higher
dielectric constant face regions of lower dielectric constant. The
amplitude of the oscillations increases by decreasing the gap size,
and the effect is progressively stronger when the contrast between
the dielectric properties of the two regions is more pronounced,
with a maximum relative change of 0.7 \% for gold-silicon, 7 \% for
silicon-air, and 65 \% for gold-air, at the closest separation of
$H=100$ nm.

In Figs. \ref{fig:norm}d-f the normal Casimir-Lifshitz forces
between the same types of structures as above are presented, for the
asymmetric case of $f=0.2$. One can see two noticeable differences
with the symmetric case. First, the oscillations are now asymmetric,
as enforced by the asymmetry of the dielectric profile, although the
asymmetry weakens as the gaps size increases and eventually
disappears---i.e. the oscillations become symmetric and
harmonic---at sufficiently large separations. This is consistent
with the picture that different harmonics of the dielectric contrast
profile in Eq. (\ref{Epp}) couple with each other via an exponential
terms that decays with the corresponding wavelengths of each
harmonic and as a result any asymmetry caused by higher harmonics
will die out at large gap sizes. The second new feature is the
significant enhancement of the amplitude of the oscillatory behavior
as a function of the lateral displacement. While it is still the
case that this amplitude increases with increasing contrast between
the dielectric properties of the two materials used in the layered
structure, the maximum relative change is 0.4 \% for gold-silicon, 6
\% for silicon-air, and 200 \% for gold-air, at the closest
separation of $H=100$ nm.

\begin{figure}
\includegraphics[width=.3\columnwidth]{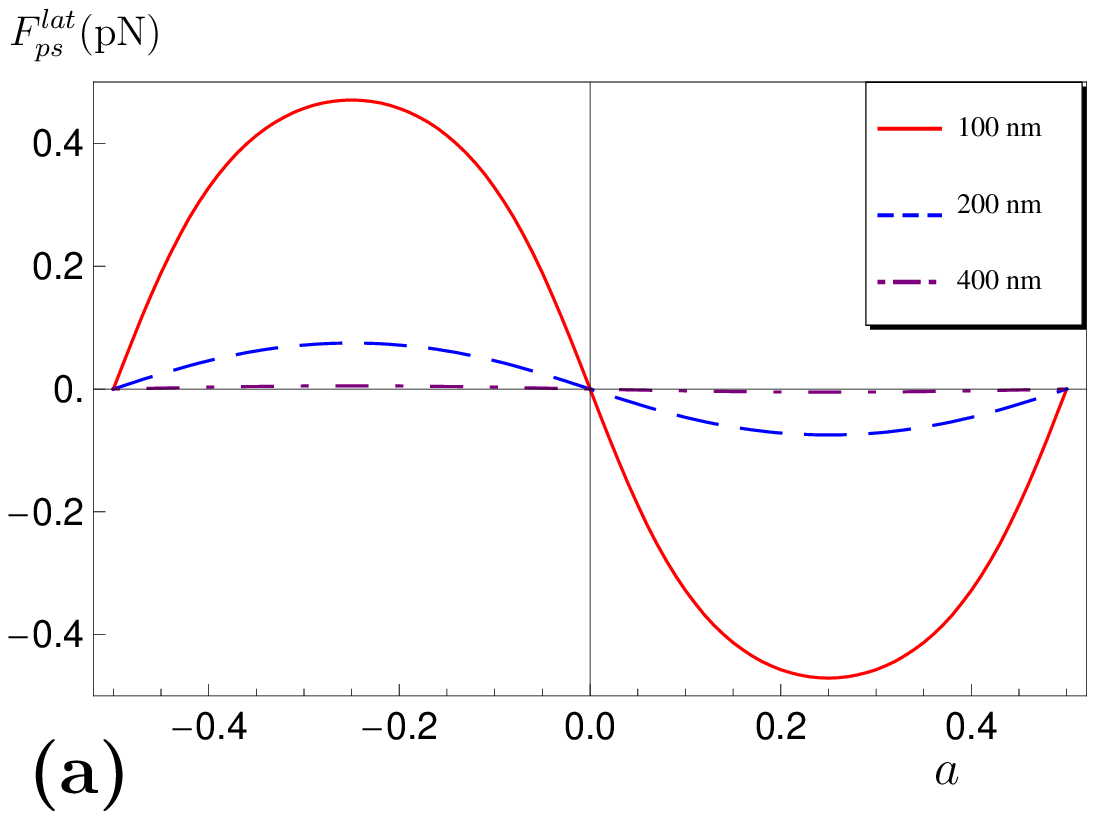}
\hspace{.5cm}
\includegraphics[width=.3\columnwidth]{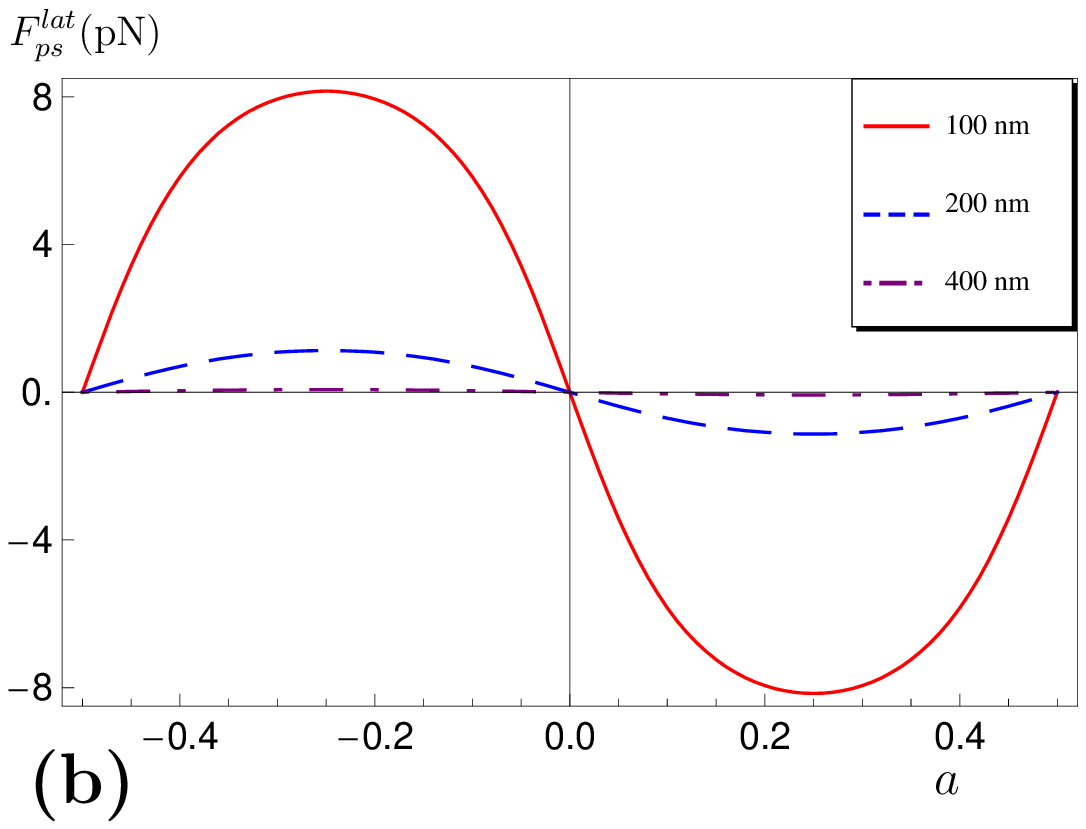}
\hspace{.5cm}
\includegraphics[width=.3\columnwidth]{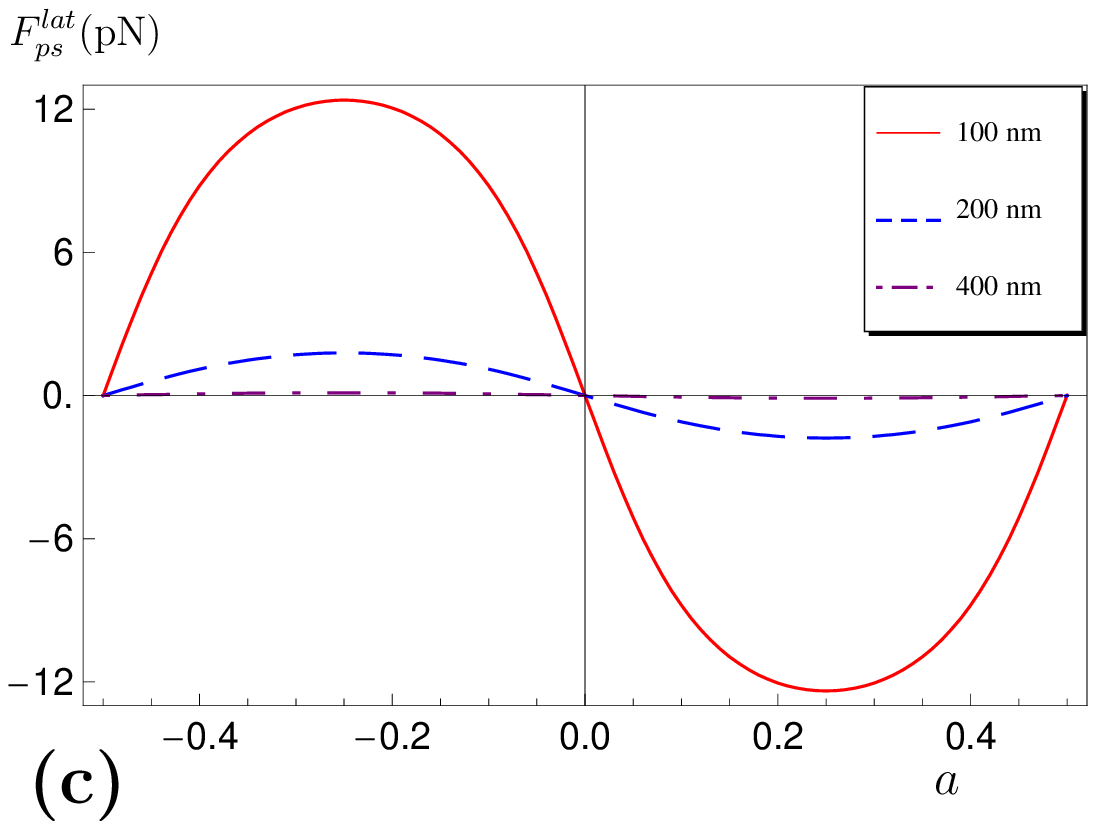}
\includegraphics[width=.3\columnwidth]{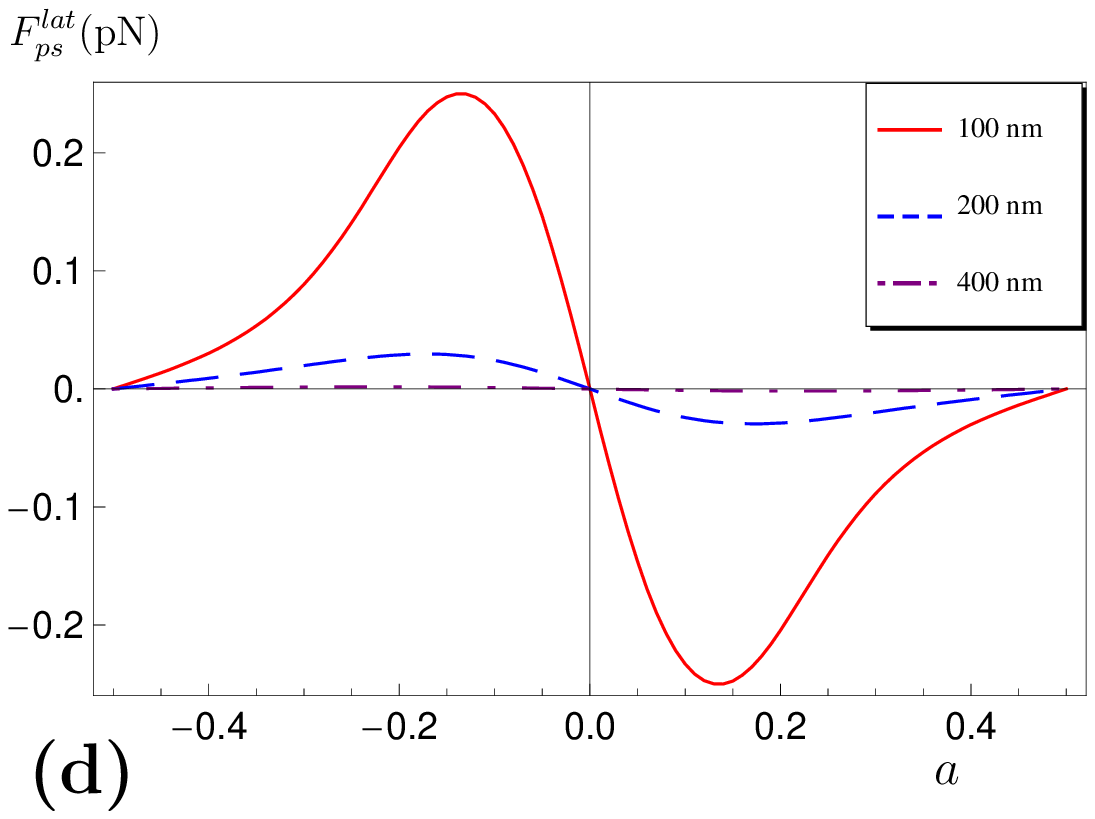}
\hspace{.5cm}
\includegraphics[width=.3\columnwidth]{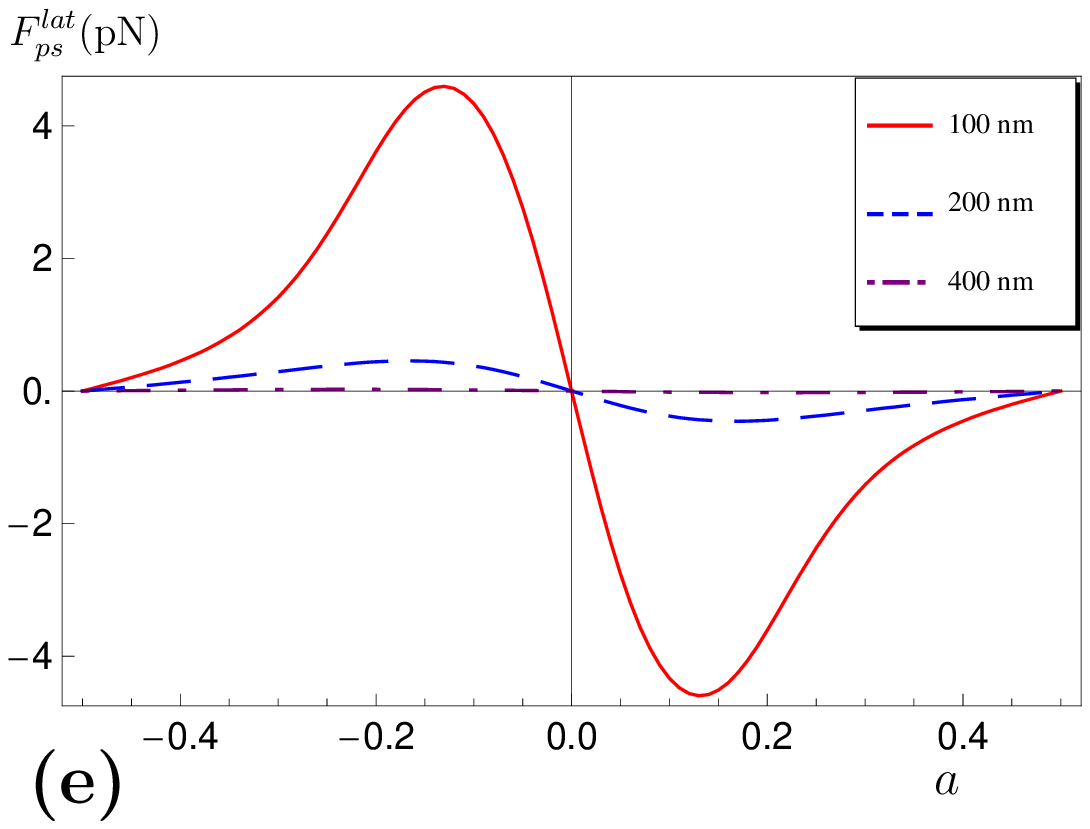}
\hspace{.5cm}
\includegraphics[width=.3\columnwidth]{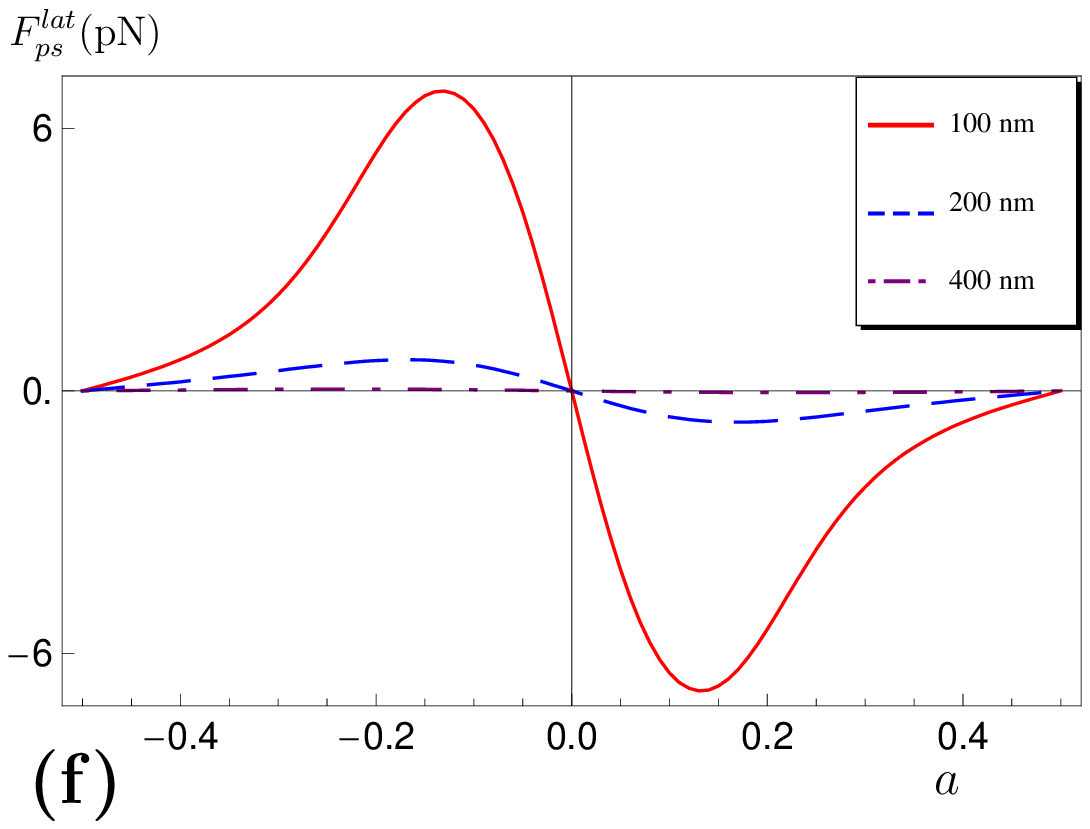}
\caption{Lateral Casimir-Lifshitz force between layered dielectric
heterostructures as shown in Fig. \ref{fig:schem} in the
plate-sphere geometry, corresponding to (a) gold-silicon, (b)
silicon-air, and (c) gold-air, with $f=0.5$, and to (d)
gold-silicon, (e) silicon-air, and (f) gold-air, with $f=0.2$. The
numerical values used in these graphs are $\lambda=1$ $\mu$m and
$R=180$ $\mu$m. Different curves correspond to different gap sizes
of $H=100$ nm, $H=200$ nm, and $H=400$ nm.} \label{fig:lat}
\end{figure}

The lateral Casimir-Lifshitz forces for the same layered structures
as above are shown in Figs. \ref{fig:lat}a-c for the symmetric case
with $f=0.5$. In this case, we have assumed $R=180$ $\mu$m and
$\lambda=1$ $\mu$m. Similar to the previous study, three different
compositions of gold-silicon, silicon-air, and gold-air are
considered each at three different gap sizes of $H=100$ nm, $H=200$
nm, and $H=400$ nm. The lateral force is found to oscillate as a
function of the lateral displacement, reminiscent of the lateral
Casimir force that is induced by geometrical corrugations
\cite{GK,lateral-exp}. The shape of the oscillatory function
approaches a sinusoidal behavior as the gap size increases,
consistent with the fact that higher harmonics do not contribute to
the force in that limit as also seen in geometrical lateral Casimir
effect \cite{Emig-exact}.  The amplitude of the oscillations
increases by decreasing the gap size as well as the contrast between
the dielectric properties of the two regions. Numerically, we find
an amplitude of 0.5 pN for gold-silicon, 8 pN for silicon-air, and
12 pN for gold-air, at the closest separation of $H=100$ nm.

Figure \ref{fig:lat}d-f show the lateral Casimir-Lifshitz forces
between the same types of structures as above, for the asymmetric
case of $f=0.2$. Similarly, the profiles of the lateral force are
noticeably asymmetric, with the asymmetry weakening as the gap size
is increased and the shape of the profile approaches that of a
sinusoidal function (single harmonic). We also see comparatively
more significant enhancement of the amplitude of the oscillatory
behavior as a function of the lateral displacement. The amplitude of
the oscillations is found as 0.3 pN for gold-silicon, 5 pN for
silicon-air, and 7 pN for gold-air, at the closest separation of
$H=100$ nm.

\section{Discussion}\label{sec:disc}

In this paper, we have proposed a mechanism by which it is possible
to create a lateral Casimir-Lifshitz force as well as controlled
modulations in the normal Casimir-Lifshitz force without geometrical
corrugations. A coupling similar to what exists in the case of
corrugated surfaces gives rise to these oscillatory forces, namely
identical modes of the dielectric patterns couple across the gap to
generate a macroscopic coherence in the fluctuations. The generic features of these oscillatory forces are very similar to those of the forces caused by corrugations; the effect is stronger and involves more harmonics at closer separations, while it weakens and only involves the lowest mode of the pattern in the dielectric contrast at larger separations.

While the difference in the dielectric properties of the materials controls the general strength of the above results, comparison between Fig. \ref{fig:norm} and \ref{fig:lat}
shows that the modulations in the normal force are more strongly affected by the contrast in the dielectric properties. The choice of air/vacuum as one component also allows us to make predictions about geometrical features with large corrugation amplitudes, which provides an approximation scheme for the non-perturbative geometrical regime.

In the present calculations we have only used the second order terms in the dielectric contrast perturbative series. Higher order terms shown in Eq. (\ref{higher}) will introduce coupling between different modes of the dielectric pattern in a systematic way, as imposed by the overall conservation of the sum of all wavevectors (momenta). While the present is aimed at showing in terms of tractable calculations, one can in principle carry out the calculation of the Casimir-Lifshitz interaction in such dielectric heterostructures using numerical diagonalization methods \cite{valery}.

Controlled interactions between dielectric heterostructures with smooth outer surfaces could be very useful in practical applications because it will help avoid the complications of bringing surfaces with geometrical protrusions close to each other while avoiding contact between them and controlling their separations. Moreover, it is much easier to pattern dielectric properties of materials in a controlled way than it is to shape them with the high precision that is needed for Casimir effect type experiments.

\acknowledgements

The authors thank the ESF Research Network CASIMIR for providing
excellent opportunities for discussion on the Casimir effect and
related topics. This work was supported by EPSRC under Grant
EP/E024076/1.

\end{document}